\documentclass[a4paper]{report}
\usepackage[utf8]{inputenc}
\usepackage[T1]{fontenc}
\usepackage{RJournal}
\usepackage{amsmath,amssymb,array}
\usepackage{booktabs}

\graphicspath{ {images/} }

\begin{document}

%% do not edit, for illustration only
\sectionhead{Contributed research article}
\volume{XX}
\volnumber{YY}
\year{20ZZ}
\month{AAAA}

%% replace RJtemplate with your article
\begin{article}
\title{\pkg{Snowboot}: Bootstrap Methods for Network Inference}
\author{by Yuzhou Chen, Yulia R. Gel, Vyacheslav Lyubchich and Kusha Nezafati}

\maketitle

\abstract{
Complex networks are used to describe a broad range of disparate social systems and natural phenomena, from power grids to customer segmentation to human brain connectome. Challenges of parametric model specification and validation inspire a search for more data-driven and flexible nonparametric approaches for inference of complex networks. In this paper we discuss methodology and  R implementation of two bootstrap procedures on random networks, that is, patchwork bootstrap of \citet{Thompson:etal:2016} and \citet{Gel:etal:2016} and vertex bootstrap of \citet{Snijders:Borgatti:1999}. To our knowledge, the new R package \CRANpkg{snowboot} is the first implementation of the vertex and patchwork bootstrap inference on networks in R. Our new package is accompanied with a detailed user's manual, and is compatible with the popular R package on network studies \CRANpkg{igraph}. We evaluate the patchwork bootstrap and vertex bootstrap with extensive simulation studies and illustrate their utility in application to analysis of real world networks. The accepted manuscript is published in the R Journal. This is an open access article under the CC BY 4.0 license (https://creativecommons.org/licenses/by/4.0/). (The R Journal (2018) 10:2, pages 95-113; doi: https://doi.org/10.32614/RJ-2018-056; URL = https://doi.org/10.32614/RJ-2018-056)
}

\section{Introduction}

Traditionally, the structural network analysis, that is, the analysis of data in the form of graphs, has been primarily approached as a descriptive task, in contrast to an inferential task, while the employed analytic tools have rooted mainly within research areas outside of ``mainstream'' statistical methodology. In the recent years, however, there has been a flare of interest in developing new statistical methods for inference on complex networks \citep[see overviews by][and references therein]{Goldenberg:etal:2010, Kolaczyk:Csardi:2014, Brugere:etal:2017}. Despite the explosive growth of statistical methods for network analysis, the developed methodology for inference on graph-structured data remains predominantly parametric and model-based. At the same time, nonparametric bootstrap and resampling appears as an appealing and flexible data-driven alternative for network inference, especially, if only a single realization of a large complex network exists. That is, we can follow the same route as the classical bootstrap of \cite{Efron:1979} which was proposed four decades ago as an alternative to conventional parametric methods for independent and identically distributed data, and later was extended to various weakly dependent space-time processes \citep{Hall:2013, Shao:Tu:1995, Chernick:2007}.

We attribute the first pioneering attempt to develop nonparametric bootstrap on networks to \citet{Snijders:Borgatti:1999}. The Snijders--Borgatti procedure, called vertex bootstrap, allows evaluating estimation uncertainty for network density and testing one- and two-sample hypotheses for densities, under the assumption that all the network information is available upfront. Vertex bootstrap is widely used in social studies and is implemented in such highly popular software for social network analysis as UCINET \citep{UCINET}. However, vertex bootstrap remains largely unknown in statistics and there still exists no implementation of this procedure in R. The next attempt to draw bootstrap inference on complex networks, with a focus on uncertainty quantification in network degree estimation, has been suggested by \citet{Thompson:etal:2016} and \citet{Gel:etal:2016}. The idea is to borrow the ``blocking'' argument, developed for bootstrapping of time series and re-tiling of spatial data, and adapt it to random networks. The resulting procedure, called patchwork bootstrap, starts from sampling a set of multiple ego networks of varying orders and forming a patch (i.e., a network block analogue), and then proceeds to resampling the data within patches. Patchwork bootstrap allows to quantify estimation uncertainties for network degree distribution and its functions, and to construct reliable and sharp confidence intervals in a fully data-driven way. Furthermore, patchwork bootstrap is applicable to ultra-sparse and only partially observable networks.

The new R package \CRANpkg{snowboot} \citep{snowboot} is the first to offer vertex and patchwork bootstrap in R. Moreover, to our knowledge, there currently exist only two more R packages implementing bootstrap analysis of networks, \CRANpkg{bootnet} and \CRANpkg{sna}. The package \CRANpkg{bootnet} assesses uncertainty in estimation of edge-weights and centrality indices but does not account for network dependence structure among vertices.
%SL: I moved the text about bootnet from above to here, because this is where we mention the packages first.
%Although the working principle of function \code{bootnet} in package \CRANpkg{bootnet} is similar to the function \code{boot\_dd} in \CRANpkg{snowboot}, the inputs are different (\code{boot\_dd} uses vertex degrees collected in a patch sample). On the one hand, labeled Snowball with Multiple Inclusions procedure can improve the running efficiency for large network; while on the other hand, we use the bootstrap network estimation methods provided by patchwork bootstrap without imposing a high risk of getting observation errors and unreliability of measurement. 
The package \CRANpkg{sna} implements parametric bootstrapping of edges to generate new random graphs that can be further used for hypothesis testing of matching between the observed network and randomized baseline network as well as canonical correlation coefficients. This paper aims to further fill the gap and showcase utility of nonparametric resampling for network inference, with a particular focus on vertex and patchwork bootstraps. \CRANpkg{snowboot} imports the R packages
\CRANpkg{graphics}, %base R
\CRANpkg{igraph} \citep{igraph}, 
\CRANpkg{parallel}, %base R
\CRANpkg{Rcpp} \citep{Rcpp}, 
\CRANpkg{Rdpack} \citep{Rdpack}, 
\CRANpkg{stats}, and  %base R
\CRANpkg{VGAM} \citep{VGAM}.

The paper is organized as follows. In the next section, we discuss methodology and implementation of vertex and patchwork bootstraps. In the simulation studies, we evaluate the implemented bootstrap methods in application to synthetic data. Our case studies illustrate application of the bootstrap to analysis of airline networks, power grids and the David Copperfield network. The paper is concluded by a discussion.

\section{Bootstrap algorithms on networks}
\label{sec:BootAlg}

\subsection{Patchwork bootstrap}

While the first results on sampling on networks go back to 1960s \citep[see, e.g.,][]{Goodman:1961, Frank:1969, Granovetter:1976} and while nowadays there exist numerous graph sampling procedures \citep[see overviews by][and references therein]{Carrington:2011, Ahmed:etal:2014, Kolaczyk:2009, simpson2015catching,  Zhang:etal:2015}, still surprisingly little is known on how to reliably and efficiently quantify sampling uncertainties, without imposing typically unverifiable model specification constraints. In this section, we discuss the new method of patchwork sampling and bootstrap  (based on algorithms of \citealp{Thompson:etal:2016} and \citealp{Gel:etal:2016}) that enables us to quantify sampling estimation uncertainties for network degree distribution and its functions, while using only a small proportion of network information.

\begin{figure}[h]
  \centering
	\includegraphics[width=0.35\textwidth, viewport= 240 53 555 517, clip]{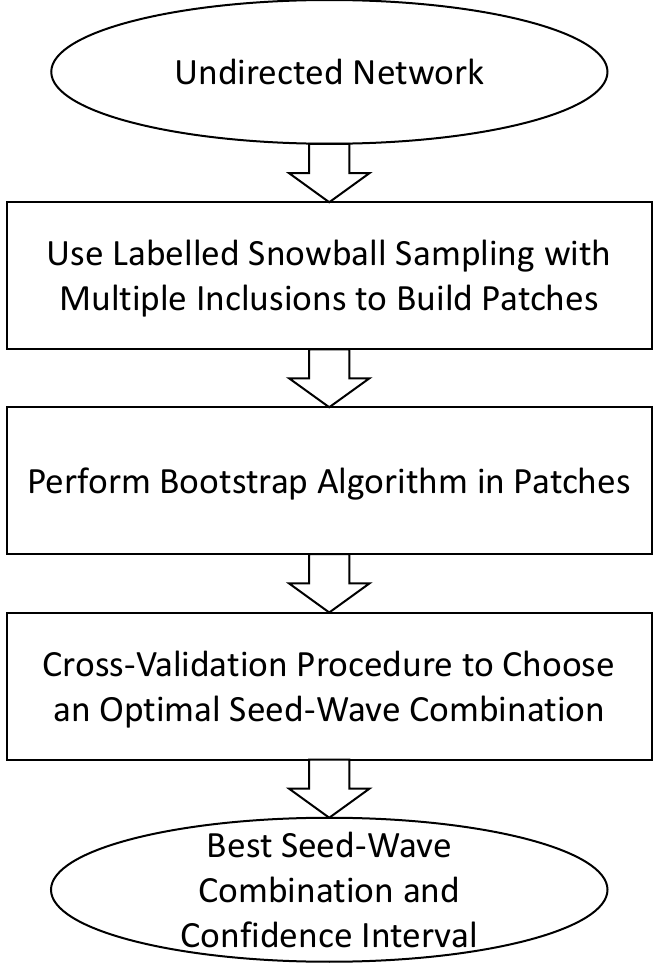}
	\caption{The workflow of the patchwork bootstrap.}
	\label{fig:diag}
\end{figure}

\subsubsection{Assumptions}\label{assum}

Consider an undirected random graph $G=(V, E)$ with a set of vertices, $V(G)$, and a set of edges, $E(G)$. The order and size of $G$ are defined as the number of vertices and edges in $G$, i.e., $|V(G)|$ and $|E(G)|$, respectively. We assume that $G$ %is loopless,
has no self-loops, i.e., $u\neq v$ for any edge $e_{uv} \in E$. The degree of a vertex $v$ is the number of edges incident to $v$.
We denote the probability that a randomly selected vertex has a degree $k$ by $f(k)$, the degree distribution of $G$ by $F=\{f(k), k\geq 0\}$, and the mean degree of $G$ by $\mu(G)$. 

Let graph $G$ represent some  hypothetical ``true'' random graph of interest that is never fully observed, and its degree distribution $F$ with finite mean and its order are unknown. Instead, we observe a random graph $G_{n}$ of order $n$ with a degree distribution $F_{n}=\{f_{n}(k), k\geq 0\}$. Let $N^{(n)}_k$ be the number of vertices with a degree $k$ in $G_n$. %The
Observed graph $G_n$ can be viewed as a realization of $G$ in a sense that as $n\to\infty$, $N^{(n)}_k/n\to f(k)$ in probability (empirical distribution $F_n$ converges in probability to $F$) and joint degree distribution of $G_{n}$ approaches that of $G$ %\cite[For the related type of convergence in configuration models see][and references therein.]{Britton:etal:2006, vanderHofstad:2014}
(see \citealp{Britton:etal:2006, vanderHofstad:2014} and references therein).

Furthermore, we assume that $G$ is {\it involution invariant} \citep{Lovasz:2012, Orbanz:Roy:2013, Crane:2018}.
%that is, from the vantage point of any randomly selected vertex, the rest of the connected network is probabilistically the same.
Note that involution invariance is linked to unimodularity~\citep{Aldous:Lyons:2007}.
That is, let us select a vertex $v$, $v \in V$ and perform a single step random walk from $v$ to one of its neighbors $u$, $u\in V$;
then the distribution of the neighborhood 
of $v$ will be the same as the distribution of the neighborhood of $u$. I.e., from the vantage point of any randomly selected vertex, the rest of the connected network is probabilistically the same.
The property of involution invariance can be viewed as a network analogue of stationarity of stochastic processes~\citep{Orbanz:Roy:2013, Crane:2018}. Indeed, as per~\cite{Aldous:Lyons:2007},
unimodularity, or involution invariance, 
relates to ``statistical homogeneity'' or ``spatial stationarity'' of a network.
In turn, stationarity is typically an essential condition for consistency of block bootstrap for space and time dependent data, thus, again linking our bootstrap procedure with the ``blocking'' argument. 
However, similarly to strong stationarity in time series analysis, involution invariance is not a formally {\it verifiable} condition. %and typically some alternative, weaker assumptions are used instead. 
In practical terms of network analysis, the proposed patchwork bootstrap 
is applicable to networks that are believed to be ``homogenous'', i.e. their distributional properties are the same across the whole network, which includes, for example, but not limited to exchangeable graphs~\citep{Crane:2018}.
%YGL: Here I comment out the next two sentences for the arxiv version
%In turn, the patchwork bootstrap will not be, for example, applicable to networks with community structures.Note that in contrast to exchangeable networks which are dense, involution invariance (and the proposed patchwork bootstrap) is also applicable to sparse networks~\citep{Orbanz:2017}. 

To the best of our knowledge, there currently exists no competing bootstrap procedure for quantifying uncertainty in estimators of network degree distribution.
The subsampling procedure of~\cite{Bhattacharyya:Bickel:2015} focuses on assessment of uncertainty in motif, or subgraph counts in dense exchangeable networks and is not feasible for inference on degree distribution. The method of~\cite{Ali:etal:2016} also targets inference
on counts of small sub-graphs and is based on the notion of dependency graphs under the network exchangeability framework. 
The vertex bootstrap approach of~\cite{Snijders:Borgatti:1999}, implemented in our R package \CRANpkg{snowboot}, does not impose density limitations but
assumes availability of the whole network upfront and is applicable to only small networks. In turn, \cite{Lusseau_etal_2008} and \cite{Epskamp:etal:2018} propose bootstrap for edge-weights and centrality indices without accounting for network dependence structure among vertices. 

The workflow of the proposed patchwork bootstrap consists of the three key steps, that is, Labeled Snowball Sampling with Multiple Inclusions, Resampling, and Cross-Validation (see Figure~\ref{fig:diag}).

\subsubsection{Labeled snowball sampling with multiple inclusions}

\begin{figure}[h]
	\centering
	\includegraphics[width=0.9\textwidth]{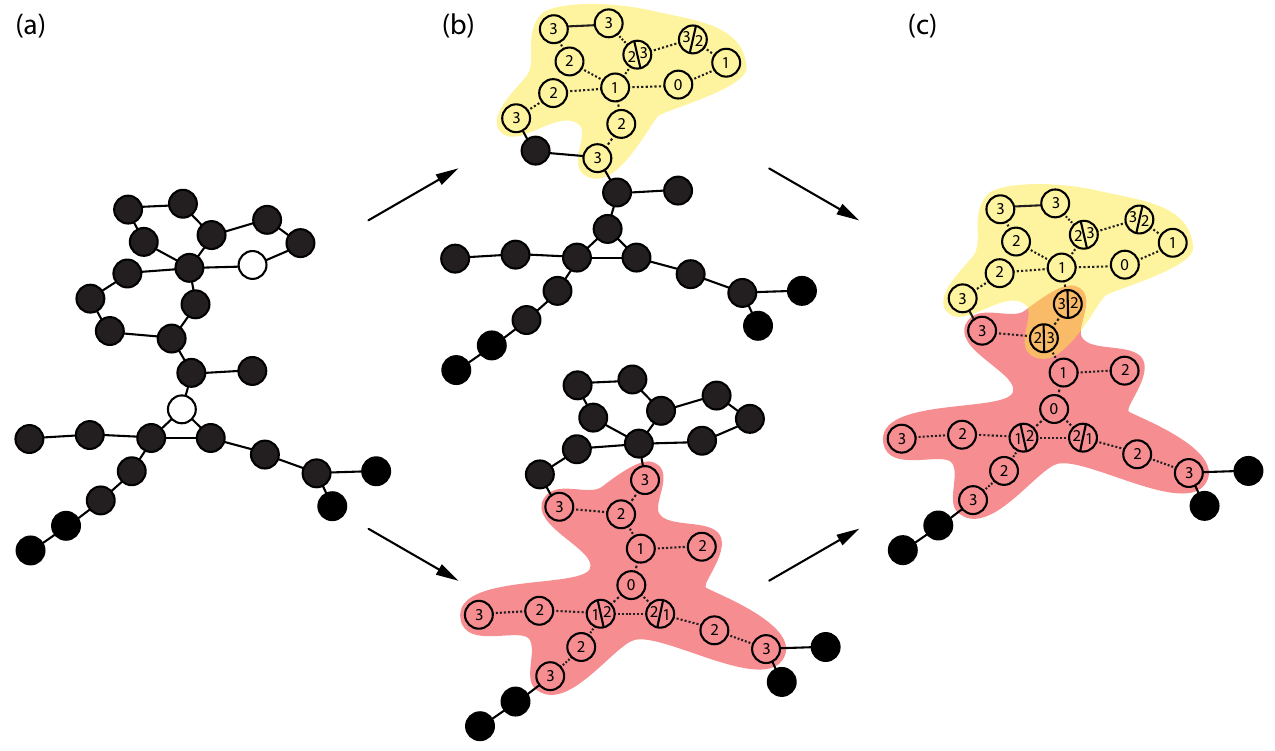}
	\caption{(a) Two seeds sampled from a network; (b) Two LSMIs grown independently, up to the third wave; (c) A patch sample of two seeds and three waves is obtained by joining the two LSMIs together. The waves are denoted with numbers from 0 (seed) to 3 (third wave) in each LSMI.}
	\label{fig:LSMI}
\end{figure}

The central element of our technique for network sampling and inference is \dfn{patch}, which is a structured sample of vertices and edges joining them. Patch sampling is performed in the following three steps:
\begin{enumerate}
\item Sample randomly without replacement several vertices from a network (Figure~\ref{fig:LSMI}(a)).
\item Construct a Labeled Snowball with Multiple Inclusions (LSMI) independently around each vertex (Figure~\ref{fig:LSMI}(b)):
\begin{enumerate}
  \item Label each of the sampled vertices as a \dfn{seed}.
  \item Select adjacent vertices by following the edges emanating from the seed and label them as \dfn{the first wave of non-seeds}.
  \item Select and label neighborhoods of second and higher orders by following the edges emanating from the first wave of non-seeds. A vertex with degree $k>1$ is included into the sample as many times as it is selected by following the previously unused edges (multiple inclusions are allowed).
\end{enumerate}
\item Join all LSMIs into one patch (Figure~\ref{fig:LSMI}(c)).
\end{enumerate}

\begin{figure}
  \centering
  \includegraphics[width=0.95\textwidth, viewport = 91 430 497 751, clip]{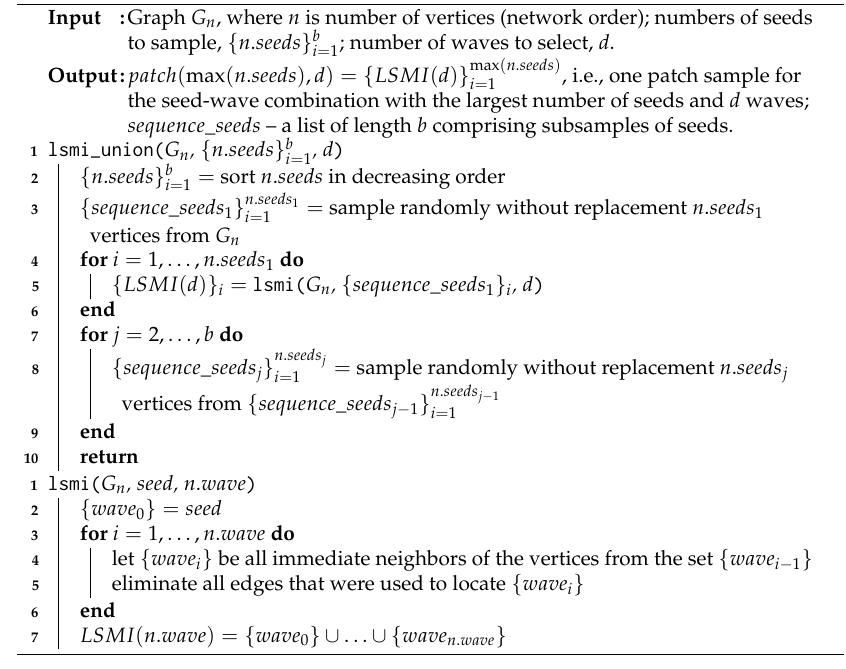}
  \caption{Obtaining a set of patches with $b$ different number of seeds, and waves from 1 to $d$.}
  \label{alg:LSMI}
\end{figure}

Detailed steps of the patch formation are given in Figure~\ref{alg:LSMI}, which is a modified version of respective algorithms by \citet{Thompson:etal:2016} and \citet{Gel:etal:2016}. The advantage of the algorithm in Figure~\ref{alg:LSMI} is that we explore patches with a smaller number of seeds by taking a subset from the seeds we have sampled, rather than by sampling new seeds from a network. This further improves computational efficiency of the algorithm and minimizes the amount of information obtained from a network.

To demonstrate the algorithm, we use one of the artificial networks stored in the R package \CRANpkg{snowboot}:

\begin{example}
> library(snowboot)
> net <- artificial_networks[[1]]
\end{example}

Function \code{lsmi} allows us to create a patch with seeds being randomly sampled or pre-specified. For example, a patch with two random seeds and one wave of neighbors around them:
\begin{example}
> set.seed(1)
> lsmi(net, n.seed = 2, n.wave = 1)
  [[1]]
  [[1]][[1]]
  [1] 532
  [[1]][[2]]
  [1] 524 763

  [[2]]
  [[2]][[1]]
  [1] 744
  [[2]][[2]]
  [1] 145 858
\end{example}
The output is structured as a list of length equal to the number of seeds sampled. Each element of this list is a list itself, where first element contains the seed ID; second element contains IDs of vertices in the first wave, and so on. This structure allows us to keep track of neighborhoods around each seed separately, including the labels -- waves in which the vertices appear. The above output shows that two seeds were sampled, with IDs \code{532} and \code{744}. The first-order neighbors of vertex \code{532} are \code{524} and \code{763}; of vertex \code{744} -- vertices \code{145} and \code{858}.

%contains IDs of the sampled vertices (\code{\$seeds} are IDs of seeds only, \code{\$sampleN} are IDs of all sampled vertices, including the seeds) and is used by other functions in the \CRANpkg{snowboot} package, e.g., to obtain the empirical degree and bootstrap degree distributions.

Alternatively, we can specify particular seeds and select the neighborhoods around them:
\begin{example}
> lsmi(net, seeds = c(532, 744), n.wave = 1)
\end{example}
This option can be used to study specific vertices in a network \citep[e.g., select the neighborhood around Erd\"os in the Erd\"os collaboration network of mathematical scientists;][]{Thompson:etal:2016}.

The next function, \code{lsmi\_union}, records information about patches obtained by subsetting the originally sampled seeds (Figure~\ref{alg:LSMI}):
\begin{example}
> patches <- lsmi_union(net, n.seeds = c(2, 5, 10), n.wave = 2)
> ls(patches)
  [1] "lsmi_big"       "sequence_seeds"
> patches$sequence_seeds
  [[1]]
  [1]  124  352  403  411 1146 1255 1319 1795 1816 1886

  [[2]]
  [1]  411 1146 1319 1816 1886

  [[3]]
  [1] 1146 1886
\end{example}
The output is a list of two elements: a patch with the biggest number of seeds and waves from those specified (in the example above, \code{patches\$lsmi\_big} is a patch with 10 seeds and 2 waves around each seed) and a list of seeds and their subsamples (in the example above, those are vectors of lengths 10, 5, and 2 with IDs of the vertices) for constructing smaller patches.

Current version of the LSMI algorithm (Figure~\ref{alg:LSMI}) pays special attention to counting the edges, thus, to precise recording of the vertices' degrees. In each patch, estimates of the probabilities for a vertex to have a certain degree $k$ ($k>0$) are obtained with a modified Horvitz--Thompson estimator, whereas $\hat f(0)$, the probability of vertices with zero degree, is approximated by the proportion of seeds with zero degree, $\hat{p}_0$. These estimates can be employed to calculate functions of the network degree distribution, e.g., mean degree $\mu$ \citep{Thompson:etal:2016, Gel:etal:2016}:
\begin{eqnarray} \label{eq:Estimates}
\hat{f}(0)&=&\hat{p}_0 = \frac{|\{d_s=0\}|}{|\{d_s\}|} ,\nonumber  \\
\hat{f}(k)&=&\frac{|\{d_s=k\}|+(1-\hat{p}_0)\hat{\mu}_s|\{d_{ns}=k\}|k^{-1}}{|\{d_s\}|+\hat{\mu}_s\sum_{k\geqslant 1}|\{d_{ns}=k\}|k^{-1}}, \\
\hat{\mu}(G) &=& \sum_{k \geqslant 0}{k \hat{f}(k)}, \nonumber
\end{eqnarray}
where ${d_{s}}$ are the degrees of the sampled seeds; ${d_{ns}}$ are the degrees of non-seeds;  %$\hat{p}_0$ is the proportion of vertices with zero degree in the set $\{d_s\}$; 
$\left| \cdot \right |$ denotes cardinality of a set; $\hat{\mu}_s$ is the estimated mean degree based on $\{d_s\}$:
\begin{equation*} \label{eq:Ek}
\hat{\mu}_s=\sum_{k\geqslant 0}k\frac{|\{d_{s}=k\}|}{|\{d_{s}\}|}.
\end{equation*}
Since the probability of non-seed vertices to be included into an LSMI is proportional to their degree, the estimates $\hat{f}(k)$ in~(\ref{eq:Estimates}) use information from non-seeds downweighted by $k^{-1}$.

Using the \code{lsmi\_dd} function, calculate $\hat{f}(k)$ from the patch we obtained earlier:
\begin{example}
> empdd <- lsmi_dd(patches$lsmi_big, net)
> empdd$mu
  [1] 2.40574
> empdd$fk
            0           1           2           3           4           5 
  0.000000000 0.369724254 0.248171075 0.207653348 0.082301632 0.040517727 
            6           7           8           9 
  0.004220597 0.025323579 0.016460326 0.005627462
\end{example}
The output is an object of class \code{'snowboot'} containing the estimates of mean degree and degree distribution. The length of the latter vector depends on what was the maximal degree ($\max(k)$) of the vertices included into the current patch. In the example above, $\max(k) = 9$.

\subsubsection{Resampling procedure}

To quantify uncertainty associated with the sample estimates~(\ref{eq:Estimates}) of network statistics, we apply a weighted bootstrap procedure. The probability of each non-seed vertex to appear in a bootstrap sample is assigned a weight of $d_{ns}^{-1}$, i.e., reciprocal of the vertex's degree. We obtain $B$ bootstrap samples and compute network statistic on each of them to approximate the distribution of the sample estimates. Each combination of number of seeds and waves gives different estimates~(\ref{eq:Estimates}), hence, the bootstrap procedure is applied separately to patches of different seed-wave combinations. The bootstrap counterparts of the estimates~(\ref{eq:Estimates}) are as follows:
\begin{eqnarray} \label{eq:BootEstimates}
\hat{f}^*(0)&=& \hat{p}^*_0 = \frac{|\{d^*_s=0\}|}{|\{d^*_s\}|}, \nonumber\\
\hat{f}^*(k)&=&\frac{ |\{d^*_s=k\}|+(1-\hat{p}^*_0) |\{d^*_{ns}=k\}|}
{ |\{d^*_s\}| + |\{d^*_{ns}\}| },\\
\hat{\mu}^*(G) &=& \sum_{k \geqslant 0}{k \hat{f}^*(k)}, \nonumber
\end{eqnarray}
where $\{d^*_s\}$ and  $\{d^*_{ns}\}$ are the respective sets of bootstrapped seeds and non-seeds.

The empirical bootstrap degree distribution can be obtained using the \code{boot\_dd} function from the \CRANpkg{snowboot} package:
\begin{example}
> B <- 50
> bootdd <- boot_dd(empdd, B)
\end{example}

A part of the output is a $(1 + \max(k)) \times B$ matrix of bootstrap estimates $\hat{f}^*(k)$, where $k=0,1,\ldots,\max(k)$, and a list of length $B$ with $\hat{\mu}^*$:
\begin{example}
> dim(bootdd$fkb)
  [1] 10 50
> length(bootdd$mub)
  [1] 50
\end{example}

The bootstrap degree distributions are used to quantify estimation uncertainty. That is, let $\eta$ be the parameter of interest (i.e., the population value, $\mu$ or $f(k)$), $\hat{\eta}^j_n$ and $\hat{\eta}_n^{j*}$ be the respective conventional and bootstrap estimators of $\eta$ based on a graph $G_n$ ($j= 1,\ldots, J$ are the different seed-wave combinations for constructing patches). 
Then the Efron's $100(1-\alpha)\%$ bootstrap confidence interval for $\eta$:
\begin{equation} \label{eq:CI}
CI^j_{percentile} = \left( \hat{\eta}_{n,\left[B\alpha/2 \right]}^{j*},\; \hat{\eta}_{n,\left[ B(1-\alpha/2) \right]}^{j*} \right),
\end{equation}
where %$j= 1,\ldots, J$ are the different seed-wave combinations, 
$B$ is the number of bootstrap samples;
$\hat{\eta}_{n,\left[B\alpha/2 \right]}^{j*}$ and $\hat{\eta}_{n,\left[ B(1-\alpha/2) \right]}^{j*}$ are the empirical quantiles from the bootstrap distribution. 

Having all bootstrapped values available from the output of \code{boot\_dd}, we can employ a variety of methods for calculating bootstrap intervals alternative to the percentile method (\ref{eq:CI}). At this time, the function \code{boot\_ci} can be switched to compute ``basic'' bootstrap intervals (see Equation~5.6 by \citealp{Davison:Hinkley:1997}):
\begin{equation}
 CI^j_{basic} = \left( 2\hat{\eta}^j_n - \hat{\eta}^{j*}_{n,[B(1-\alpha/2)]},\; 2\hat{\eta}^j_n - \hat{\eta}^{j*}_{n,[B\alpha/2]} \right).
\end{equation}

In our synthetic network example, $\hat{f}(3)=$ 0.208, and its 95\% bootstrap confidence interval (\ref{eq:CI}), based on the 50 bootstrap samples, is
\begin{example}
> CIpercentile <- boot_ci(bootdd)
> CIpercentile$fk_ci[,"3"]
       2.5%     97.5% 
  0.1286842 0.2631579
\end{example}

\begin{figure}[h]
	\centering
	\includegraphics[width=0.99\textwidth]{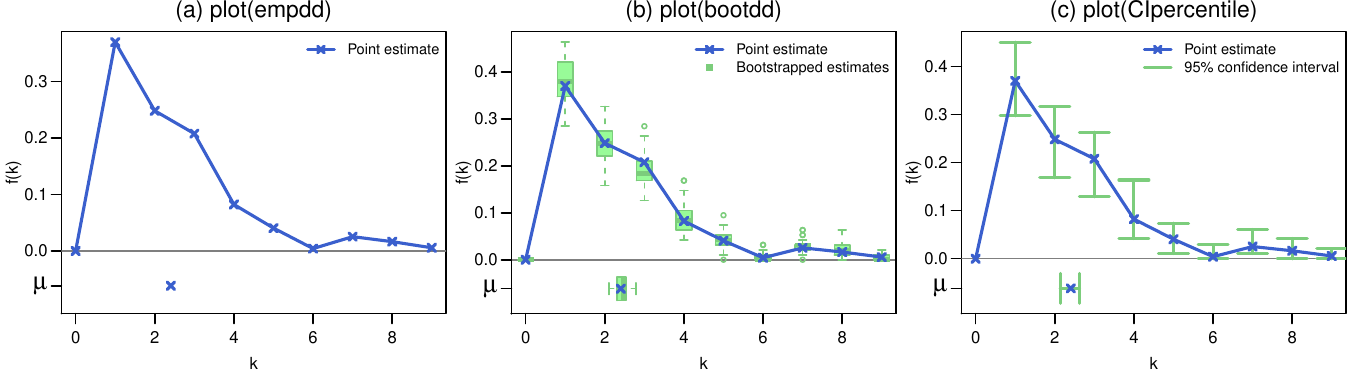}
	\caption{Results of plotting three different objects of class \code{`snowboot'}.}
	\label{fig:plotsnowboot}
\end{figure}

The outputs of the functions \code{lsmi\_dd}, \code{boot\_dd}, and \code{boot\_ci} are estimates of the network degree distribution $f(k)$ and mean degree $\mu$ based on a single patch. They are recognized as objects of class \code{'snowboot'} and can be plotted with the S3 method \code{plot} for this class (Figure~\ref{fig:plotsnowboot}). 

\subsubsection{Cross-validation with LSMI}

Growing snowball samples to higher waves is used in sampling surveys when obtaining new seeds is prohibitively expensive, for example, in  hard-to-reach and ``hidden'' subpopulations of HIV high risk individuals. Even if no more new seeds can be obtained, varying the number of seeds in a patch offers an additional flexibility, so that multiple seed-wave combinations can be evaluated and an optimal seed-wave combination is selected for a given observed network.

Given bootstrap confidence intervals from patches for each $j$th seed-wave combination ($j=1,\ldots,J$), we use Algorithm~2 of \citet{Gel:etal:2016} to decide which patch provides confidence intervals of the best coverage for our statistic $\eta$. We approximate the ground truth using proxy samples. A proxy sample is obtained by resampling (with or without replacement) vertices already used in the LSMIs, so no new information is needed from the network. From multiple such samples, proxy statistics $\hat{\eta}^{proxy}$ are estimated, and then for each $j$ a proportion of proxy statistics within the interval $CI^j$ is calculated. This proportion aims to approximate the true coverage probability of the bootstrap confidence intervals, so an interval $CI^{j_{opt}}$ with the coverage closer to the nominal level $1-\alpha$ can be selected.

Cross-validation is performed automatically by an upper-level function \code{lsmi\_cv} in \CRANpkg{snowboot} package. This function obtains multiple patches for each of the specified seed-wave combinations and runs cross-validation to select the optimal bootstrap confidence interval based on proxy values for the mean (by default, 19 proxy samples are obtained, each comprising 30 vertices):

\begin{example}
> cv <- lsmi_cv(net, n.seeds = c(10, 20, 30), n.wave = 5, B = 100)
> cv
  $`bci`
      2.5%    97.5% 
  1.749167 2.952500 
  
  $estimate
  [1] 2.387042

  $best_combination
  n.seed n.wave 
      10      2 

  $seeds
  [1]  218  323  851 1003 1039 1097 1410 1624 1723 1756
\end{example}
The output of \code{lsmi\_cv} contains the optimal seed-wave combination (\code{best\_combination}), the corresponding point estimate (\code{estimate}) and bootstrap confidence interval (\code{bci}), and, for information purposes, the IDs of the actual seeds that were used in the patch with the selected seed-wave combination (\code{seeds}). In this example, a fixed increment grid is used for the number of seeds and waves, however, a user may choose to do a stochastic search in the parameter space.

\subsubsection{Why does patchwork bootstrap work and its areas of limitations}

{\bf Bias vs. variance} Many estimators of graph totals based solely on seeds are known to be unbiased \citep{Frank:1977}. However, variance of such seed-based estimator might be high if the number of seeds is low. In turn, in many applications sampling more seeds might be prohibitively expensive, e.g., due to data privacy and cost restrictions (see overview by \citealp{Illenberger:Flotterod:2012} and references therein). Adding information from non-seeds into the degree estimator increases bias but reduces variance.
Hence, the choice of optimal number of seeds (egos) and waves of non-seeds in LSMI implies a classical bias vs.~variance trade-off, and we address it using the cross-validation procedure described above.

{\bf Network properties} Furthermore, as in many non-network settings, in order to ensure consistency of the bootstrap estimator $\hat{\eta}_n^{*}$  of the statistic of interest $\eta_n$, based on a degree distribution $F_{n}=\{f_{n}(k), k\geq 0\}$ of a random graph $G_{n}$, we typically need to ensure
that $F_{n}$ is a satisfactory approximation of $F$ as $n\to\infty$. That is, we need to ensure that the empirical degree distribution $F_{n}$ cannot be drastically different from the population degree distribution $F$ and shall approach the population degree distribution $F$ with increasing $n$, which in turn is a necessary condition for $\eta_n \to \eta$ as $n\to\infty$.
In addition, $F$ needs to satisfy some invariance properties. Those conditions give rise to the assumptions for patchwork bootstrap on p.~\pageref{assum}. Note that while nowadays there are no formal tests for assessing
involution invariance, similarly as there exists no test to assess strong stationarity in time series,
a class of involution invariant graphs includes, for example, such a large subclass as exchangeable graphs~\citep{Lovasz:2012, Orbanz:Roy:2013}. Asymptotic properties of patchwork bootstrap for $\hat{\eta}_n^{*}$ for involution invariant graphs are discussed in more details in~\citet{Gel:etal:2016}.
In addition, \citet{Thompson:etal:2016} consider the two-phase conditional inference framework to derive asymptotic properties
for patchwork bootstrap estimator of mean degree $\mu$.

{\bf Sampling design} Finally, we would like to emphasize that properties of bootstrap on networks largely depend not only on the underlying network topology but on the closely linked question on how sampling is performed. In this section we focus only on properties of a network degree distribution and argue that LSMI appears to be a suitable choice (see more discussion in \citealp{Illenberger:Flotterod:2012, Gel:etal:2016}). In turn, bootstrap and, generally, finite-sample inference for other network statistics will typically require adjustment of a sampling design~\citep{Kolaczyk:2009}.

Hence, the algorithms of LSMI sampling and estimation realized in \CRANpkg{snowboot} package target such network statistics as network degree distribution (probabilities of observing vertices with a specific degree) and its functions (e.g., mean degree and network density). In turn, the algorithm of patchwork bootstrap aims to quantify the  uncertainty of those estimates. 
%YGL: Here I comment out the next two sentences for the arxiv version
%Switching to other network statistics, such as clustering coefficient or motifs, requires implementation of different sampling and bootstrap schemes, for example, via combination of patchwork bootstrap with the algorithm of~\cite{Ali:etal:2016}
%\citep[see][for discussion on sampling and subsampling designs]{Orbanz:2017}. Hence, this extension constitutes a natural direction for future methodological research and R code implementation.

\subsection{Vertex bootstrap}
\label{vertex_bootstrap}

\begin{figure}
  \centering
  \includegraphics[width=0.9\textwidth, viewport = 91 461 503 760, clip]{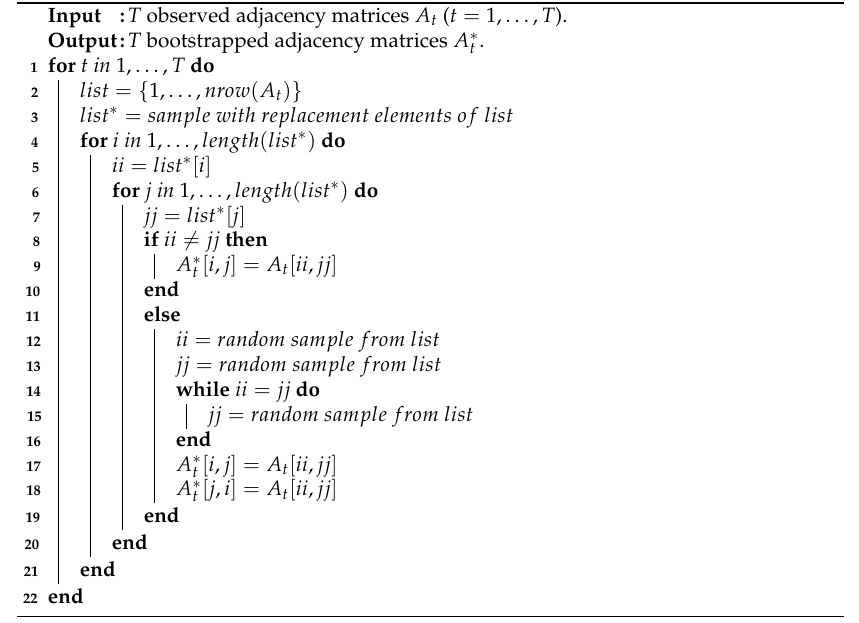}
  \caption{Vertex bootstrap.}
  \label{alg:VB}
\end{figure}

Vertex bootstrap
%(Algorithm~\ref{alg:VB}),
(Figure~\ref{alg:VB}),
or the Snijders--Borgatti procedure, is the pioneering approach to non-parametric bootstrap inference on graphs. Nevertheless, despite its high popularity in social sciences, vertex bootstrap remains largely unknown in statistics. Vertex bootstrap employs an induced graph sampling for quantification of standard errors in network density estimation and allows hypothesis testing on density of two networks. The algorithm assumes availability of the entire network data upfront as well as requires resampling of the entire data set and, hence, is limited to relatively small networks due to computational costs.
To the best of our knowledge, there exists no analysis of asymptotic properties and theoretical guarantees for the Snijders--Borgatti procedure.

We implement vertex bootstrap in the R package \CRANpkg{snowboot} as a function \code{vertboot}, which resembles calculations under the option ``Network $\rightarrow$ Compare densities'' of the UCINET software \citep{UCINET}.

%\begin{algorithm}[h!]
%\DontPrintSemicolon
%\SetKwFunction{Union}{Union}\SetKwFunction{FindCompress}{FindCompress}
%\SetKwInOut{Input}{Input}\SetKwInOut{Output}{Output}
%		\Input{$T$ observed adjacency matrices $A_{t}$ ($t=1,\ldots,T$).}
%		\Output{$T$ bootstrapped adjacency matrices $A^*_{t}$.}
%		\For{$t$ in $1,\ldots,T$}{
%		  $list = \{1,\ldots,nrow(A_{t})\}$\;
%		  $list^* = sample~with~replacement~elements~of~list$\;
%			\For {$i$ in $1,\ldots,length(list^*)$}{
%			 $ii = list^*[i]$\;
%			\For {$j$ in $1,\ldots,length(list^*)$}{
%			$jj = list^*[j]$\;
%			\If {$ii \neq jj$}{
%			  $A^*_{t}[i,j] = A_{t}[ii,jj]$\;
%			}\Else{
%			   $ii = random~sample~from~list$\;
%			   $jj = random~sample~from~list$\;
%					\While {$ii = jj$}{
%			            $jj = random~sample~from~list$\;
%			        } %EndWhile
%				$A^*_{t}[i,j] = A_{t}[ii,jj]$\;
%			  $A^*_{t}[j,i] = A_{t}[ii,jj]$\;
%			}
%		  } %end j
%			} %end i
%		} %end k
%			\caption{Vertex bootstrap}\label{alg:VB}
%\end{algorithm}

The \code{vertboot} function not only generates similar results compared with UCINET, but also returns results with higher precision and faster run times. In \CRANpkg{snowboot}, the algorithm is written in C\texttt{++}.

Another important improvement is that \code{vertboot} allows users to compare statistics of interest for multiple networks, whereas the number of different networks, $T$
%(Algorithm~\ref{alg:VB}),
(Figure~\ref{alg:VB}),
in UCINET is limited to 1 or 2. Finally, the output of \code{vertboot}
%(Algorithm~\ref{alg:VB})
(Figure~\ref{alg:VB})
is a bootstrapped network, which implies that users can analyze various network statistics besides the network density.

We demonstrate the \code{vertboot} function using prison network data. The data were collected from 67 prison inmates; each inmate could choose as few or as many ``friends'' as he desired. A direct factor analysis of these sociometric data was performed by \citet{MacRae:1960}.

\begin{itemize}
    \item Get the observed adjacency matrix for the prison network:
\begin{example}
> a <- scan("http://vlado.fmf.uni-lj.si/pub/networks/data/ucinet/prison.dat",
+            skip = 4)
> A <- matrix(a, sqrt(length(a)), byrow = TRUE)
\end{example}

    \item Apply
    %Algorithm~\ref{alg:VB}
    vertex bootstrap (Figure~\ref{alg:VB})
    $B$ times to generate $B$ bootstrapped adjacency matrices (networks):
\begin{example}
> set.seed(1)
> B <- 500
> Astar <- vertboot(A, B)
\end{example}

    \item Get the densities of $B$ bootstrapped networks:
\begin{example}
> library(igraph)
> densities <- sapply(1:B, function(x)
+    graph.density(graph_from_adjacency_matrix(Astar[[x]])))
\end{example}

    \item Estimate density of the observed network and bootstrap standard error of the estimates:
\begin{example}
> density_obs <- graph.density(graph_from_adjacency_matrix(A))
> densities_se <- sd(densities)
\end{example}

    \item Obtain a 95\% bootstrap confidence interval for the population density:
\begin{example}
> quantile(densities, c(0.025, 0.975))
        2.5%      97.5%
  0.03301673 0.05178652
\end{example}
\end{itemize}

We can use the outputs of the vertex bootstrap (Figure~\ref{alg:VB})
to build bootstrap confidence intervals for other statistics:

\begin{itemize}
    \item Mean degree:
\begin{example}
> mu_star <- sapply(1:B, function(x)
+    mean(degree(graph_from_adjacency_matrix(Astar[[x]]), mode = "in")))
> quantile(mu_star, c(0.025, 0.975))
      2.5%    97.5%
  2.179104 3.417910
\end{example}

     \item Transitivity:
\begin{example}
> net_transitivity <- sapply(1:B, function(x)
+    transitivity(graph_from_adjacency_matrix(Astar[[x]]), type = "undirected"))
> quantile(net_transitivity, c(0.025, 0.975))
        2.5%      97.5%
  0.08890533 0.21852252
\end{example}
\end{itemize}

\section{Simulation studies}
\label{sec:Simulation}

\subsection{Light- and heavy-tailed degree distributions}

\begin{figure}[h]
  \centering
  \includegraphics[width=\textwidth, viewport=0 5 625 210, clip]{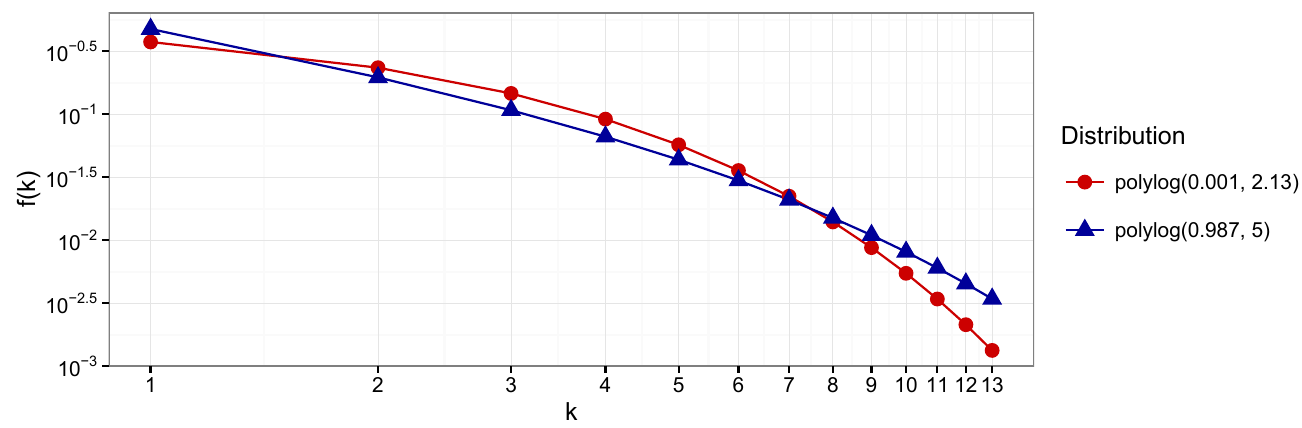}
  \caption{Polylogarithmic distributions with the same mean ($\mu=2.67$), but different thickness of tails. The graphs are log-log graphs, where both axes use a logarithmic scale.}
\label{fig:two_poly}
\end{figure}

In this section, we demonstrate the performance of the patchwork bootstrap algorithm for inference on mean degree of simulated graphs. We consider random graphs of orders 1000, 5000, and 10000 that are constructed from two polylogarithmic distributions of the same mean (population mean degree $\mu(G)=2.67$ for all simulated networks), but with different tail behavior. Particularly, polylogarithmic distribution with parameters $\delta=0.001$, $\lambda=2.13$ has a light tail, whereas polylogarithmic distribution with parameters $\delta=0.987$, $\lambda=5$ has a heavy tail (Figure~\ref{fig:two_poly}).

The results of our simulation study (Table~\ref{talbe_ver_pat_sim}) show a good performance of the patchwork bootstrap approach with automatic selection of patch sizes (i.e., seed-wave combinations) using the cross-validation procedure. Thus, the observed coverage probabilities of the confidence intervals are close to the nominal level 95\% for both light- and heavy-tailed networks. At the same time, confidence intervals in the considered light-tailed networks are approximately 25\% narrower. Time complexity of the patchwork procedure on any random graph is $\mathcal{O}([\left | d_{s}\right |\cdot \hat{\mu}_s]^{\left | d_{ns}\right |})$.

\setlength{\tabcolsep}{0.4em}
\begin{table}[h]
	\small \centering
	\begin{tabular}{cccccc}
		\toprule
		 & & &Network order $n$\\ \cmidrule{3-5}
		Distribution & Mean degree $\mu(G)$ & 1000 & 5000 & 10000\\\midrule
		polylog($\delta=0.001$, $\lambda=2.13$) & 2.67 &  97.7\% (0.794) & 96.4\% (0.812)& 97.7\% (0.805)\\
		polylog($\delta=0.987$, $\lambda=5$) & 2.67 &  98.6\% (1.051) & 98.7\% (1.078)&97.7\% (1.069)\\
		\bottomrule
	\end{tabular}
	\caption{Observed coverage probabilities of 95\% patchwork bootstrap confidence intervals for the mean degree (average interval width is in parentheses). These intervals come from the optimal seed-wave combination (one for each random graph) defined via a cross-validation over a grid of 20 combinations: waves from 1 to 5; and number of seeds 20, 30, 40, and 50.
In cross-validation, proxy mean degrees are estimated 13 times from 100 vertices sampled without replacement from the patch data. The number of bootstrap samples, $B$, is 500. The experiment uses 1000 Monte Carlo simulations, carried out as parallel processes on a distributed computing cluster.}
	\label{talbe_ver_pat_sim}
\end{table}

\subsection{Vertex removal}

In this section, we randomly remove vertices (and edges emanating from those vertices) from the simulated networks prior to applying the patchwork and vertex bootstrap methods for the inference on mean degree. This allows us to assess the methods' robustness and capabilities of providing reliable inference when part of the network information is missing. Our goal is to estimate when performance of the methods decreases significantly.

\begin{figure}[h]
	\centering
		\includegraphics[width=\textwidth]{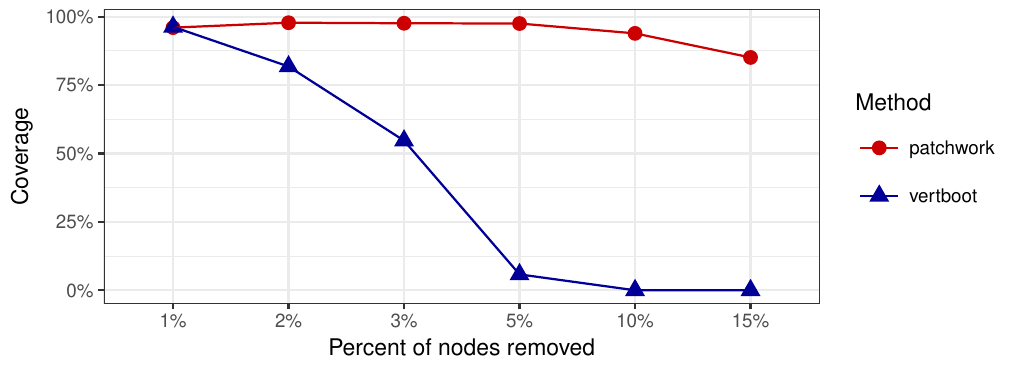}
		\caption{Observed coverage probabilities for 95\% confidence intervals for network mean degree, delivered by the two bootstrap methods when different numbers of vertices are removed from the network. Network order $n=5000$; degree distribution polylog($\delta=0.001$, $\lambda=2.13$); $B=500$; 1000 Monte Carlo simulations.}
\label{fig:node_removal_light}
\end{figure}

As Table~\ref{table_patch_sim} shows, when only 1\% of the vertices are removed, both methods deliver confidence intervals with coverage close to the declared 95\% level. When 2\% of network vertices are removed, coverage probabilities of vertex bootstrap intervals decline to 81.8\% and 89.8\% for light- and heavy-tailed networks, respectively. The performance of vertex bootstrap further rapidly declines as more vertices are removed, and the decline is more severe for the light-tailed polylogarithmic distribution. Remarkably, patchwork bootstrap performs consistently well on both distributions (observed coverage probabilities are close to the nominal level), even when 5\% of the vertices are removed (Table~\ref{table_patch_sim}). Using the example of the light-tailed polylogarithmic distribution, we show that coverage of vertex bootstrap declines to zero when 10\% or 15\% of vertices are removed, while respective coverage of the patchwork bootstrap intervals is 93.9\% and 85.1\%, respectively (Figure~\ref{fig:node_removal_light}).

\begin{table}[h]
	\small\centering
	\begin{tabular}{lcccccc}
		\toprule
		 &Mean&& \multicolumn{4}{c}{Number of vertices removed} \\ \cmidrule{4-7}
		Distribution&  degree $\mu(G)$& Method & 1\% & 2\% & 3\% & 5\%\\
		\midrule
		%\multirow{6}{*}{polylog($\delta=0.001$, $\lambda=2.13$)} & \multirow{6}{*}{2.67}&\multirow{3}{*}{Patchwork}
        polylog($\delta=0.001$, $\lambda=2.13$) & 2.67 &Patchwork &96.0\%& 97.8\%&97.6\%&97.5\%\\
		&& & (0.807) & (0.801)&(0.806)&(0.795)\\
		&& & [0.187] & [0.194]&[0.191]&[0.196]\\ \cmidrule{3-7}
		%&&\multirow{3}{*}{Vertex bootstrap}
        && Vertex bootstrap & 96.3\%& 81.8\%&54.7\%&5.8\%\\
		&& & (0.170)& (0.170)&(0.169)&(0.169)\\
		&& & [0.090]& [0.091]&[0.090]&[0.090]\\ 	\midrule
		%\multirow{6}{*}{polylog($\delta=0.987$, $\lambda=5$)} & \multirow{6}{*}{2.67}&\multirow{3}{*}{Patchwork}
        polylog($\delta=0.987$, $\lambda=5$) & 2.67 & Patchwork &99.4\%& 99.0\%&98.9\%&99.4\%\\
		&& & (1.072) & (1.062)&(1.062)&(1.032)\\
		&& & [0.230] & [0.239]&[0.249]&[0.237]\\ \cmidrule{3-7}
		%&&\multirow{3}{*}{Vertex bootstrap}
        && Vertex bootstrap & 96.3\%& 89.8\%&74.0\%&25.7\%\\
		&& & (0.212)& (0.211)&(0.211)&(0.209)\\
		&& & [0.113]& [0.113]&[0.113]&[0.112]\\
		\bottomrule
	\end{tabular}
    \caption{Observed coverage probabilities of 95\% nonparametric bootstrap confidence intervals. Average interval width is in parentheses, standard errors are in square brackets. Network order $n=5000$; number of bootstrap samples $B=500$; 1000 Monte Carlo simulations. For the patchwork method, $J=20$ seed-wave combinations: number of seeds 20, 30, 40, and 50; number of neighbors from 1 to 5.}	
\label{table_patch_sim}
\end{table}

Hence, the patchwork bootstrap approach is a competitive alternative when analyzing large complex networks, both in terms of computational speed and reliability of inference when a part of the network information is missing. Moreover, the patchwork method is both computationally efficient and information-greedy, i.e., only a small proportion of the network data is required, the procedure subsets the sampled seeds to consider patches of different sizes and re-uses information from the patches in the cross-validation procedure. In contrast, the vertex bootstrap employs information of the whole target network so that its time complexity is $\mathcal{O}(n^{2})$. At the same time, for small networks with all the network information being available upfront, the vertex bootstrap is the preferred method as it provides noticeably sharper confidence intervals under the same level of calibration.

\section{Case studies}
\label{sec:CaseStudies}

In this section, we illustrate utility of the \CRANpkg{snowboot} package for analysis of airline networks, power grids, and the David Copperfield network.

\subsection{The David Copperfield network}

\begin{figure}[h]
	\centering
	\includegraphics[width=0.5\textwidth, viewport=17 135 580 705, clip]{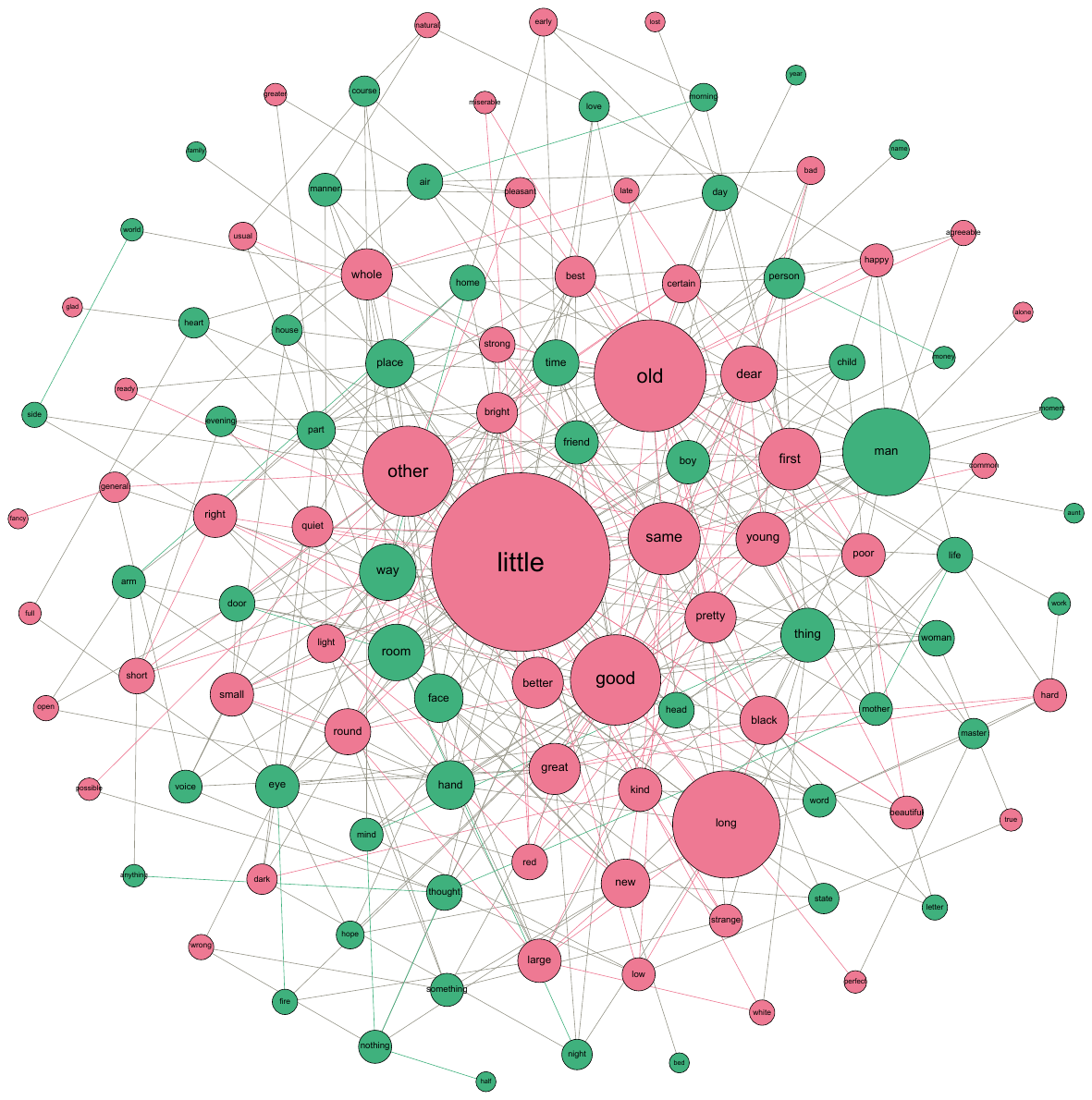}
	\caption{Lexical network: graph of nouns and adjectives found in the novel David Copperfield.}
	\label{fig:david_graph}
\end{figure}

We start from a smaller network, namely, the David Copperfield network collected by \cite{Newman:2006}. It examines the lexicon of Charles Dickens's classic 19th century novel. The network vertices are common nouns and adjectives; undirected edges connect adjacent words (Figure~\ref{fig:david_graph}).

The number of vertices and edges in David Copperfield network are 112 and 425, respectively. This is a relatively small network, so the vertex bootstrap algorithm is suitable for analysis. Some basic network statistics of the David Copperfield network are presented in Table~\ref{tab:Copp}.

\begin{table}[h]
	\centering
\begin{tabular}{*{5}{c}}
\toprule
	Order& Density & Number of edges & Mean degree & Clustering coefficient  \\ \midrule
	112 & 0.068 & 425 & 7.589 & 0.157   \\
\bottomrule
\end{tabular}
\caption{Parameters of the David Copperfield network.}
\label{tab:Copp}
\end{table}

Perform the vertex bootstrap in the following steps:
\begin{itemize}
	\item Load the network data and obtain an adjacency matrix:
\begin{example}
> library(igraph)
> graph_david <- read.graph(
+    "http://networkdata.ics.uci.edu/data/adjnoun/adjnoun.gml", format = "gml")
> A <- as.matrix(as_adjacency_matrix(graph_david))
\end{example}

  \item Use vertex bootstrap (Figure~\ref{alg:VB}) to obtain bootstrapped adjacency matrices:
\begin{example}
> library(snowboot)
> B <- 500
> set.seed(1)
> Astar <- vertboot(A, B)
\end{example}

	\item Use these bootstrapped networks to calculate 95\% bootstrap confidence intervals for the density and bootstrap standard error:
\begin{example}
> boot_density <- sapply(1:B, function(x)
+    graph.density(graph_from_adjacency_matrix(Astar[[x]])))
> CIvertboot <- quantile(boot_density, c(0.025, 0.975))
> CIvertboot
        2.5%      97.5%
  0.05473174 0.08579271
> bootstrap_standard_error <- sd(boot_density)
> bootstrap_standard_error
  [1] 0.007681788
\end{example}

That is, we find that the resulting 95\% confidence interval from the vertex bootstrap is (0.055, 0.086); the interval contains the observed density of 0.068. Now let us apply the patchwork bootstrap to the David Copperfield network.

		\item Use the patchwork bootstrap to calculate 95\% bootstrap confidence intervals for the density:
\begin{example}
> set.seed(5)
> igraph_david <- igraph_to_network(graph_david)
> CIpatchwork <- lsmi_cv(igraph_david, n.seeds = c(3:5), n.wave = 1, B = B)
> CIpatchwork$bci/(igraph_david$n - 1)
        2.5%      97.5% 
  0.04305405 0.08957658
\end{example}
\end{itemize}

We find that the 95\% confidence interval from the patchwork bootstrap also contains the observed density of 0.068. However, the patchwork bootstrap CI is wider than the vertex bootstrap CI (i.e., 0.047 for patchwork vs. 0.031 for vertex procedure). At the same time, the patchwork bootstrap used only about 31.4\% of the available vertices.

\subsection{Larger networks}

\subsubsection{Flight networks of the airline alliances}

With constantly increasing costs of operation and rigid legal restrictions on ownership of national flag carriers, no single airline can provide a comprehensive global network, which is vital to its success. An urgent need for expansion propels airlines to effectively cooperate on multiple levels: from interlining (combining flights from different airlines in one travel itinerary) to joint frequent flyer programs, to sharing revenue, costs, and benefits \citep{IATA:2011}.

Nowadays, three major passenger airline alliances (Star Alliance, Oneworld, and SkyTeam) share more than 70\% of the world market. There exists a constant, fierce competition in customer acquisition and retention among these three alliances, and one of its key factors is claimed to be a route map (number of served destinations and better connections). Traditional indicators, such as total passenger enplanements or number of aircraft movements, fail to capture the competitiveness of airline networks \citep{Burghouwt:Redondi:2013}. However, for a fixed network structure, e.g., with hubs and spokes, dense air flight networks (with a high mean degree) are more convenient, since they further minimize the required number of connections.  The International Air Transport Association (IATA) study of European Union countries showed that a 10\% rise of the number of destinations served and/or the frequency of service can increase long-run gross domestic product by 1.1\% \citep{IATA:2006}.

We then evaluate densities of the three flight networks (Star Alliance, SkyTeam, and Oneworld) to assess which alliance offers the most convenient travel network to its passengers.  Since flight connections are easier if a transfer occurs within one airport, we focus on airport networks.  This approach treats all airports as separate vertices even some of them have the same service area (for example, Toronto Pearson International Airport and Billy Bishop Toronto City Airport).

\begin{figure}[h!]
\centering
%\begin{subfigure}{0.5\textwidth}
%    \centering
(a) Star Alliance
    \includegraphics[width=0.99\textwidth, viewport=62 142 1246 695, clip]{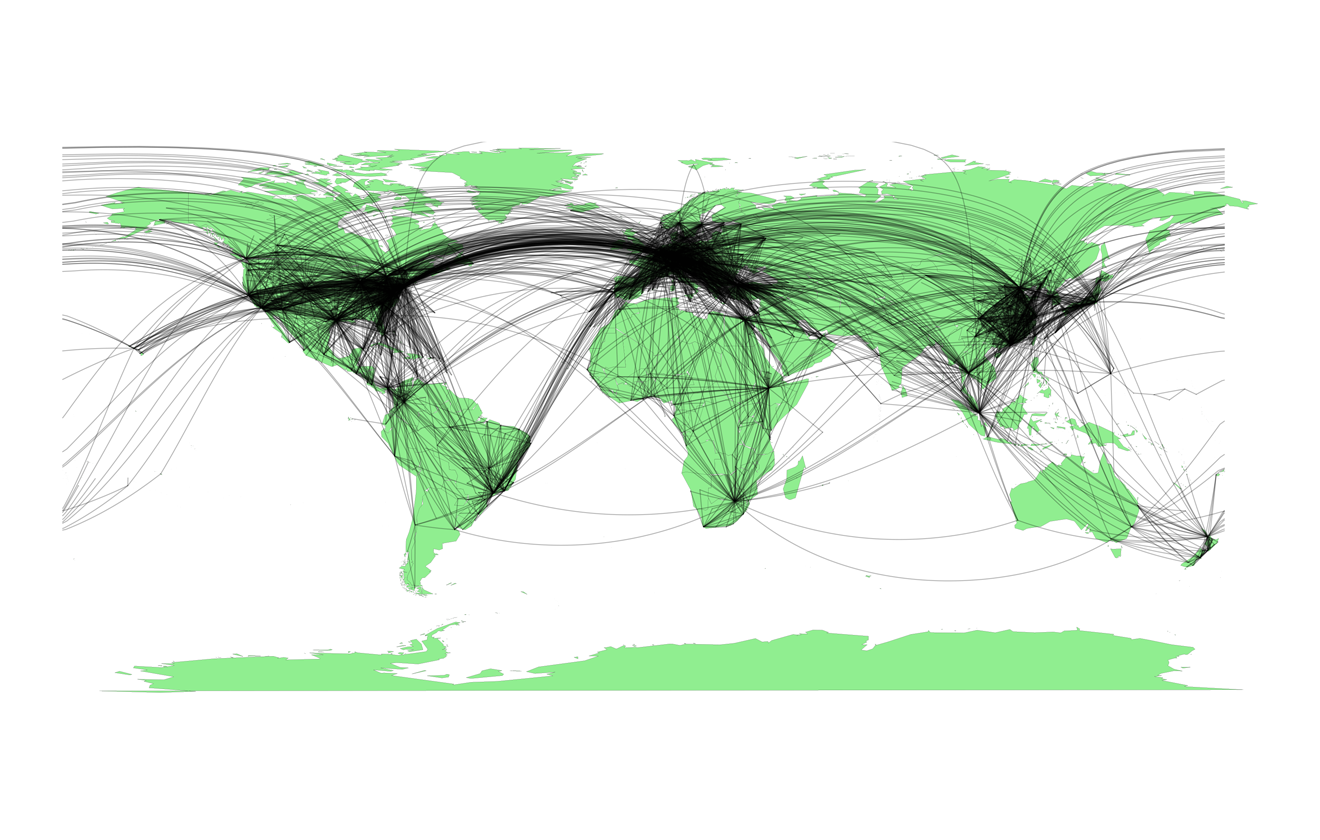}
%    \caption{Star Alliance: $n$=1,289; $\hat{d}(G_n)$=$0.00621$}
%    \label{net_StarAlliance}
%\end{subfigure}%
%\begin{subfigure}{0.5\textwidth}
%    \centering
(b) SkyTeam
    \includegraphics[width=0.99\textwidth, viewport=62 142 1246 695, clip]{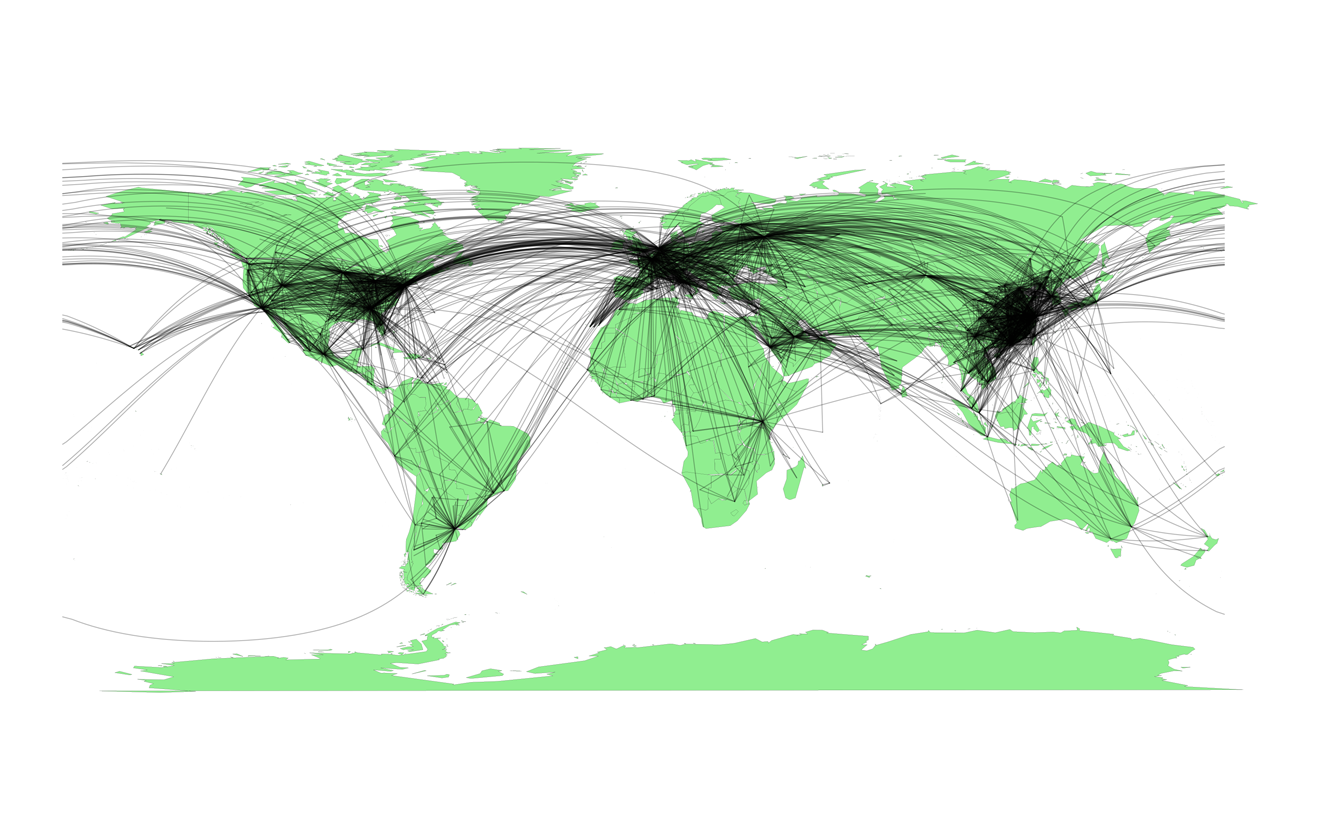}
%    \caption{SkyTeam: $n$=1,040; $\hat{d}(G_n)$=$0.00736$}
%    \label{net_SkyTeam}
%\end{subfigure}
%\begin{subfigure}{0.99\textwidth}
%    \centering
(c) Oneworld
    \includegraphics[width=0.99\textwidth, viewport=62 142 1246 695, clip]{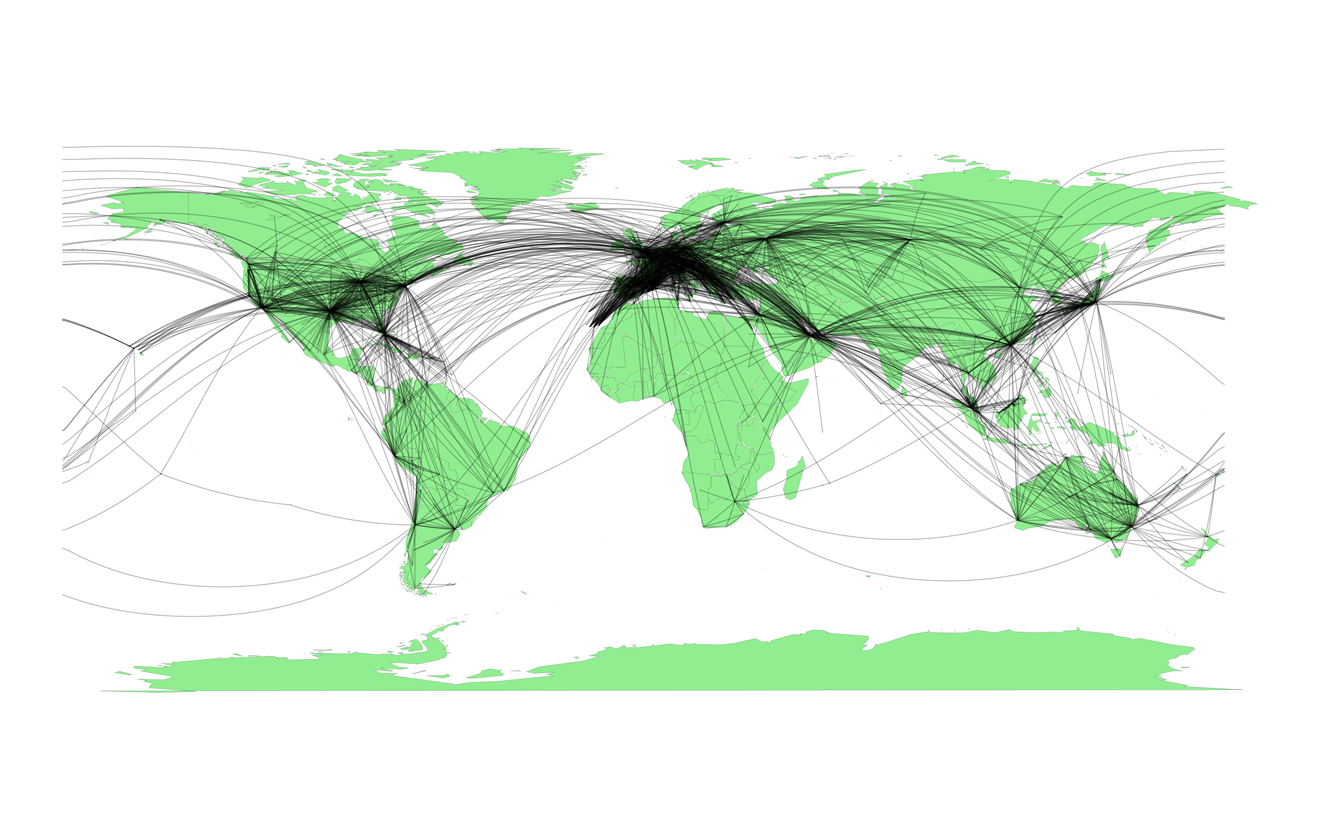}
%    \caption{Oneworld: $n$=914; $\hat{d}(G_n)$=$0.00655$}
%    \label{net_OneWorld}
%\end{subfigure}
\caption{Airport networks of the airline alliances.}
\label{fig:alliances}
\end{figure}

To construct the networks, we obtained the crowdsourced air flights data from OpenFlights.org on October 23, 2013.  At the time of analysis, Star Alliance, Oneworld, and SkyTeam included 28, 13, and 19 members, respectively.  We used the airline codes to select all flights for each airline alliance, but removed self-loops and the cases with missing airport identifications (up to 0.4\% of the records).  To eliminate the repeated entries, including the codeshare flights within each alliance, we kept only unique links between airports.  About 11.0\% of all remaining vertices had different in-degree and out-degree, but we neglected the directions for simplicity.  Thus, for each airline alliance, we obtained a network where the vertices stand for airports, and each unweighted undirected edge represents existing flight connection at least in one direction (Figure~\ref{fig:alliances}).

Table~\ref{table_air} reports the networks orders $n$ and network densities $\hat{d}(G_n)$ estimated on all observed data along with the results of the patchwork bootstrap. The network order (i.e., number of served airports) is close to 1,000 for all three alliances. (Given a relatively high order of the airline networks, we do not consider the vertex bootstrap in this case study.) The optimal seed-wave combinations suggested by the cross-validation procedure differ among the networks: 20 seeds and 1 wave for SkyTeam, 30 seeds and 1 wave for Star Alliance, and 40 seeds and 1 wave for Oneworld. The width of the confidence interval for Star Alliance and Oneworld is only 0.00296 and 0.00321, respectively, compared with 0.00497 for SkyTeam. Since all three confidence intervals overlap with each other, we conclude that currently there exists no significant difference in flight connections offered by the three major airline alliances. Thus, the loyalty of frequent fliers and acquisition of new customers are likely to be attributed to other factors, such as customer service, loyalty program benefits, ticket prices, and availability of flights in particular regions (e.g., Figure~\ref{fig:alliances} %net_OneWorld
shows that Oneworld network provides almost no service in African airports).

\begin{table}[h]
\centering
\begin{tabular}{rrrrrrr}
\toprule
%&&&\multicolumn{2}{r}{\makecell[r]{Optimal\\combination}}& \multicolumn{2}{r}{\makecell[r]{95\% confidence bounds\\for the density $d(G)$}}\\ \cmidrule{4-7}
&&&\multicolumn{2}{r}{Optimal}& \multicolumn{2}{r}{95\% confidence bounds}\\
&&&\multicolumn{2}{r}{combination}& \multicolumn{2}{r}{for the density $d(G)$}\\ \cmidrule{4-7}
Network&$n$&$\hat{d}(G_n)$& Seeds& Waves&\hspace{2em}Lower& Upper\\ \midrule
Star Alliance&	1289&	0.00621&	30&	1& 0.00492& 0.00788\\ %9.38-5.94 = 3.44
SkyTeam	     &  1040&	0.00736&	20&	1& 0.00561& 0.01058\\ %10.41-4.06 = 6.35
Oneworld     &	  914&	0.00655&	40&	1& 0.00533& 0.00854\\ %8.17-4.23 = 3.94
\bottomrule
\end{tabular}
\caption{The 95\% bootstrap confidence intervals for the density of airline alliance networks, replicating connections of the airports (vertices) with the flights of member airlines (edges).  Considered 12 seed-wave combinations: waves from 1 to 3, seeds 20, 30, 40, and 50.  Number of bootstrap resamples is 500 per each combination. Cross-validation is based on a random selection of 100 vertices 10 times.}
\label{table_air}
\end{table}

\subsubsection{United States and Germany power grids}

The power grid serves as the backbone of critical infrastructure sector and is essential for today's society as an enabling infrastructure. A combination of three substations, i.e., generator, transmission, and distribution, connected by high voltage transmission lines, provide the United States and Germany with electrical power people so heavily rely on. As with many other large scale infrastructures, the power grid serves users who may notice its presence and realize its importance only when the system fails in some way. One of the main issues with the system is that failures or disruptive events like hurricanes, earthquakes, and attacks, can cause cascading failures in a power grid. As the power grids increase in size and complexity, it is of a paramount importance to study their vulnerability.

\begin{figure}[h!]
\centering
\includegraphics[width=0.95\textwidth]{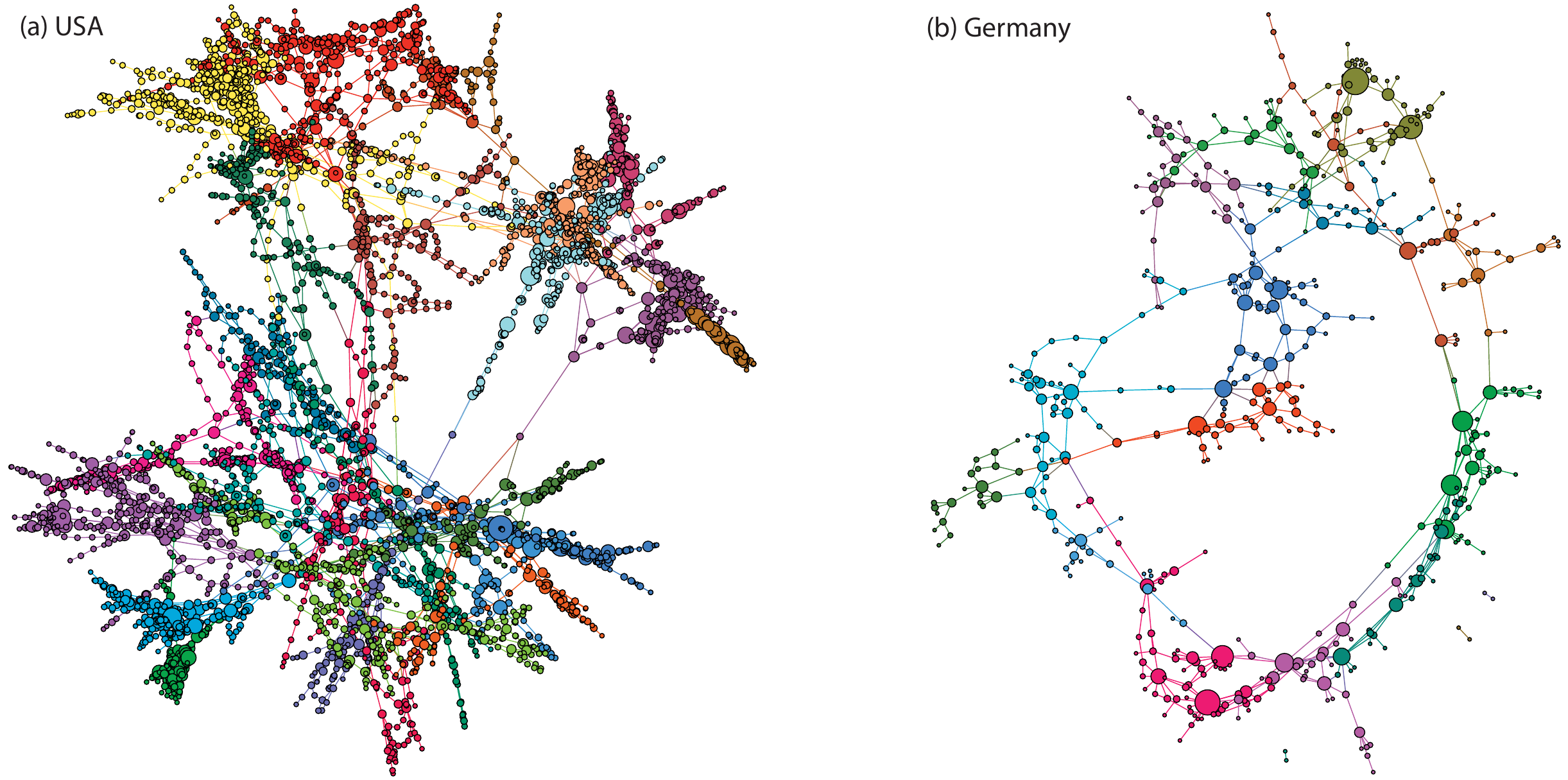}
\caption{Power grids of the western states of the United States and Germany. The colors show network modules; connections between the vertices within modules are denser than between vertices from different modules.}
\label{fig_power_grid}
\end{figure}

To better understand the effects of power system failures, the power grids can be analyzed from the perspective of random networks. In this case study, we consider two power grids that are represented as undirected networks, that is, the power grids of the western states of the United States \citep{Watts:Strogatz:1998} and Germany \citep{SciGRIDv0.2}. Each edge represents a power supply line and vertex is a generator, a transformer, or a substation (Figure~\ref{fig_power_grid}).

Centrality statistics are one of the most widely explored attributes of a power grid network. Some studies focus on the relationship between various centrality statistics and resilience of the power grid networks \citep{pagani:etal:2013}. Another potential indicator of power system robustness and resilience is network density~\citep{Cuadra_et_al2015}. In addition, \citet{Sole_et_al2008} and \citet{RosasCasals_CorominasMurtra2009} propose to use a characteristic parameter $\gamma$, based on fitting an exponential distribution to an empirical cumulative distribution of each grid as a classifier of grid fragility. That is, the network is robust if $\gamma < 1.5$ and is fragile otherwise. In this study, we would like to examine the difference in fragility properties of the power grids in Germany and the western states of the United States, in terms of the $\gamma$ parameter and the proportion of distribution stations in high-voltage networks. Given a relatively high order of the networks, the vertex bootstrap is not feasible, and we apply the patchwork bootstrap to compare the two power grids (see Table~\ref{table_Power}). We find that both power grids deliver the characteristic parameter $\hat{\gamma}$ of higher than 1.5, that is, 2.09 and 2.62 for the US and Germany, respectively, and hence both grids shall be classified as fragile. However, the respective 90\% patchwork bootstrap confidence intervals do not overlap, and we can conclude that the US power grid in the western states tends to be less fragile than the German power grid. Remarkably, the 90\% confidence intervals for densities of the two networks also do not overlap, hence, indicating that there likely exists a significant difference in connectivity of the two power grid networks.

While it is pre-mature to conclude that the Western US power grid system is more robust than the German power system due to its higher sparsity, especially given the lack of a uniformly accepted notion of power system fragility and robustness~\citep{pagani:etal:2013, Cuadra_et_al2015,Dey:etal:2017:GlobalSIP, Islambekov:etal:2018}, it is reasonable to conclude that the two systems exhibit significant differences in their structure. In turn, the bootstrap methodology provides a route how power grids and their network properties can be systematically evaluated and compared in a framework of statistical hypothesis testing.

\setlength{\tabcolsep}{0.4em}
\begin{table}[h!]
	\centering
	\begin{tabular}{rrrrrrrr}
		\toprule
		&&&&\multicolumn{4}{c}{90\% confidence bounds}\\
    &&&&\multicolumn{2}{c}{$\gamma$}&\multicolumn{2}{c}{$d(G)$}\\\cmidrule{5-8}
		Power grid network&$n$&$\hat{\gamma}$&$\hat{d}(G_n)$&\hspace{2em}Lower& Upper&\hspace{2em}Lower& Upper\\ \midrule
		United States&	4941&	2.08943&0.00054& 1.75241& 2.12021&0.00048&0.00056\\
		Germany&523&2.62055&0.00642&2.20408&2.78395&0.00586&0.00692\\
		%Germany power grid\star&511&2.40&2.10&3.15\\
		\bottomrule
	\end{tabular}
	\caption{The 90\% bootstrap confidence intervals for the fragility parameter $\gamma$ and density $d(G)$ of two power grid networks.  Considered 20 seed-wave combinations: waves from 1 to 5, seeds 20, 30, 40, and 50.  Number of bootstrap resamples is 500 per each combination. Cross-validation is based on a random selection of 100 vertices 13 times.}
\label{table_Power}
\end{table}

\section{Conclusion}
\label{sec:Conclusion}

In this paper we discuss utility and implementation of the two bootstrap methods for nonparametric inference on complex networks, that is, patchwork bootstrap of \citet{Thompson:etal:2016} and \citet{Gel:etal:2016}  and vertex bootstrap of \citet{Snijders:Borgatti:1999}. We primarily focus on developing inference on network degree distribution and its functions, e.g., mean and density. Furthermore, we perceive the observed network data as a single realization of some ``true'' unobserved network, and our target is to draw statistical inference in a model-free data-driven way, given only this single network realization. While there is an ever increasing interest in nonparametric inference on complex networks and despite the fact that the vertex bootstrap has been implemented in UCINET software for social network analysis for more than a decade, to our knowledge, there exists no single implementation of bootstrap methods on graphs in R. Our new package \CRANpkg{snowboot} fills this gap and offers a flexible data-driven alternative for parametric analysis of complex networks. Furthermore, \CRANpkg{snowboot} is fully compatible with \CRANpkg{igraph}, and provides a number of options, such as Labeled Snowball Sampling with Multiple Inclusions and cross-validation on graphs -- the functionality of standalone interest for graph mining and network analysis.

\section{Acknowledgements}

The authors thank Steve P. Borgatti for the help with the vertex bootstrap and Marti Rosas-Casals for the help with  interpretation of $\gamma$ fragility parameter in the power grid analysis. This work was partially supported by grants from the National Science Foundation (NSF) of the United States, IIS 1633331/1633355,  DMS 1736368 and ECCS 1824710.

\bibliography{chen_gel_lyubchich_nezafati_arxiv}

\address{Yuzhou Chen\\
  Department of Statistical Science\\
  Southern Methodist University\\
  P.O. Box 750332\\
  Dallas, Texas, 75275\\ %-0332
  USA\\
  E-mail: \email{yuzhouc@smu.edu}
}

\address{Yulia R. Gel, Kusha Nezafati\\
  Department of Mathematical Sciences\\
  University of Texas at Dallas\\
  800 West Campbell Road\\
  Richardson, Texas, 75080\\
  USA\\
  E-mail: \email{ygl@utdallas.edu}, \email{kusha.nezafati@utdallas.edu}}

\address{Vyacheslav Lyubchich\\
  Chesapeake Biological Laboratory\\
  University of Maryland Center for Environmental Science\\
  146 Williams Street / P.O. Box 38 \\
  Solomons, Maryland, 20688\\
  USA\\
  E-mail: \email{lyubchich@umces.edu}}
  
\end{article}

\end{document}